\documentclass[aps,superscriptaddress,10pt,nofootinbib,notitlepage,twocolumn,prd]{revtex4-1}

\usepackage{hyperref}
\usepackage{xcolor}
\usepackage{graphicx}
\usepackage{physics}
\usepackage{bm}
\usepackage{amssymb}

\renewcommand{\a}{a}

\newcommand{\be}[1]{\begin{equation} #1 \end{equation}}

\newcommand{\SM}[1]{\text{\sc sm}}
\newcommand{\KKLT}[1]{\text{$\mathrm{KKLT}$}}
\newcommand{\LVS}[1]{\text{$\mathrm{LVS}$}}

\def\be{\begin{equation}}
\def\ee{\end{equation}}

\def\a{\alpha}
\def\vp{\varphi}
\def\K{K{\"a}hler }
   \newcommand{\rf}[1]{(\ref{#1})}

\begin{document}

\title{The BAO-CMB Tension and Implications for Inflation}

\author{Elisa G. M. Ferreira}
\affiliation{Kavli IPMU (WPI), UTIAS, The University of Tokyo, 5-1-5 Kashiwanoha, Kashiwa, Chiba 277-8583, Japan}
\affiliation{Center for Data-Driven Discovery, Kavli IPMU (WPI), UTIAS, The University of Tokyo, Kashiwa, Chiba 277-8583, Japan}

\author{Evan McDonough}
\affiliation{Department of Physics, University of Winnipeg, Winnipeg MB, R3B 2E9, Canada}

\author{Lennart Balkenhol}
\affiliation{Sorbonne Université, CNRS, UMR 7095, Institut d’Astrophysique de Paris, 98 bis bd Arago, 75014 Paris, France}

\author{Renata Kallosh}
\affiliation{Stanford Institute for Theoretical Physics, Stanford, CA 94305, USA}

\author{Lloyd Knox}
\affiliation{Department of Physics and Astronomy, University of California, Davis, CA, 95616 USA}

\author{Andrei Linde}
\affiliation{Stanford Institute for Theoretical Physics, Stanford, CA 94305, USA}

\begin{abstract}
The scalar spectral index $n_s$ is a powerful test of inflationary models. The tightest constraint on $n_s$ to date derives from the combination of cosmic microwave background (CMB) data with baryon acoustic oscillation (BAO) data. The resulting $n_s$ constraint is shifted significantly upward relative to the constraint from CMB alone, with the consequence that previously preferred inflationary models are seemingly disfavored by $\gtrsim 2 \sigma$. Here we show that this shift in $n_s$ is the combined effect of a degeneracy between $n_s$ and BAO parameters exhibited by CMB data and the tension between CMB datasets and DESI BAO data under the assumption of the standard cosmological model. Given the crucial role of $n_s$ in discriminating between inflationary models, we urge caution in interpreting CMB+BAO constraints on $n_s$ until the BAO-CMB tension is resolved.
\end{abstract}

\maketitle

\tableofcontents{}

\section{Introduction}
\label{sec:intro}
\parskip 5pt

Cosmic inflation is a leading candidate to describe the very early universe. 
Inflation has passed many tests \cite{Kallosh:2025ijd}, starting with the observation of the acoustic peaks of the cosmic microwave background (CMB), which excluded then-competitor cosmic strings as the origin of structure 
\cite{Magueijo:1996px,Pen:1997ae,Allen:1997ag,Knox:2000dc,Dodelson:2003ip,Dvorkin:2011aj,Urrestilla:2011gr}, and subsequently by the WMAP \cite{WMAP:2012nax} and {\it Planck} \cite{Planck:2018jri} experiments' inference of a primordial power spectrum of scalar perturbations that is adiabatic, Gaussian, and nearly-scale invariant, in agreement with the simplest models of inflation.

While ongoing and upcoming experiments \cite[e.g.][]{BICEP:2021xfz,SPT-3G:2025vtb,Ade:2018sbj, LiteBIRD:2022cnt,CMB-S4:2016ple}
pursue, as a primary science goal, constraints on primordial gravitational waves via their imprint in the B-mode polarization of the CMB (the ``Holy Grail'' \cite{Abazajian:2013vfg} of inflation\footnote{See \cite{Brandenberger:2011eq} for an alternative view.}), the spectral index $n_s$ remains a powerful discriminator of inflationary models.
For example, many models of inflation, such as racetrack inflation, D3/D7 inflation, and inflection point inflation, that were compatible with WMAP can be excluded by {\it Planck} on the basis of $n_s$ alone \cite{Kallosh:2019eeu,Kallosh:2025ijd}.

 Recent works by the ACT~\cite{ACT:2025fju} and SPT~\cite{SPT-3G:2025bzu} collaborations report joint constraints on $n_s$ from CMB and BAO data that are 
 shifted upward from those reported by Planck.
 The combined data set of {\it Planck} and ACT (including CMB lensing), and BAO from the Dark Energy Spectroscopic Instrument (DESI)  DR1, gives the ``P-ACT-LB" constraint $n_s=0.9739\pm 0.0034$ \cite{ACT:2025fju}, which is nearly identical to the constraint from the combination of SPT with Planck and ACT (``SPA'') and DESI DR2 BAO, the latter given by $n_s = 0.9728\pm 0.0027$ \cite{SPT-3G:2025bzu}, both in $\Lambda$CDM. If confirmed, this will have important implications for inflation: the {\it Planck} preferred inflationary models of Starobinsky \cite{Starobinsky:1980te}, Higgs inflation \cite{1989PhRvD..40.1753S,Bezrukov:2007ep}, and many (though not all) $\alpha$-attractors \cite{Kallosh:2013yoa}, are disfavored at $\gtrsim 2\sigma$. This has led to a flurry of activity to reconcile these inflationary models with the latest CMB data or to extend the already existing models matching the latest CMB data \cite{Kallosh:2025rni,Maity:2025czp,Heidarian:2025drk,Frolovsky:2025iao,Chakraborty:2025jof,Lizarraga:2025aiw,Cacciapaglia:2025xqd,Addazi:2025qra,Kim:2025dyi,Pallis:2025epn,Odintsov:2025wai,Gialamas:2025kef,McDonald:2025odl,Yin:2025rrs,Aoki:2025wld,Antoniadis:2025pfa,Haque:2025uis,He:2025bli,Salvio:2025izr,Pallis:2025nrv,Haque:2025uga,Hai:2025wvs,Saini:2025jlc,Kohri:2025lau,Liu:2025qca,Drees:2025ngb,Zharov:2025evb,Yogesh:2025wak,Byrnes:2025kit,Okada:2025lpl,Byrnes:2025kit,Yogesh:2025wak,Peng:2025bws,McDonald:2025tfp,Choudhury:2025vso,Mohammadi:2025gbu}.

To understand this sudden interest, one may take a look at Fig. 2 in \cite{Chang:2022tzj}, which shows only three targets of the B-mode search by CMB-S4 and LiteBIRD: the Starobinsky model, Higgs inflation, and $\alpha$-attractors. It does not show any viable targets with $n_s\gtrsim 0.97$.

However, as detailed by the SPT collaboration in Ref.~\cite{SPT-3G:2025bzu}, DESI DR2 data is in tension with CMB data under the assumption of $\Lambda$CDM.  The tension with ACT is $3.1\sigma$, and when ACT is combined with SPT-3G, the tension increases to $3.7\sigma$. These quantifications of tension are based on differences between the BAO parameters, $r_dh$ and $\Omega_m$, as inferred from the DESI and CMB datasets. 
Note that, with the assumption of $\Lambda$CDM, BAO data can be losslessly compressed to constraints
on these two parameter combinations. Since these parameters can also be inferred from CMB data, this two-dimensional
parameter space is a natural one for assessing consistency, under $\Lambda$CDM, of CMB and BAO datasets. We henceforth refer to the tension in $r_d h$ and $\Omega_m$ as the {\it BAO-CMB tension}.

The BAO-CMB tension can be interpreted as an indication of physics beyond $\Lambda$CDM or else of some unknown systematic errors. Indeed, the BAO-CMB tension drives a $2-3\sigma$ preference for beyond-$\Lambda$CDM models \cite{SPT-3G:2025bzu}.
The tension also draws into question the validity of combining the CMB and DESI BAO data sets in the context of $\Lambda$CDM\footnote{See e.g.~Ref.~\cite{LuisBernal:2018drn} for a discussion of the risks of combining inconsistent data sets.}.

In this paper, we revisit CMB constraints on $n_s$ in light of the BAO-CMB tension. 
We demonstrate that the fit of $\Lambda$CDM to CMB data exhibits a high degree of correlation between $n_s$ and the BAO parameters $r_dh$ and $\Omega_m$. This correlation, and the tension in $r_dh$ and $\Omega_m$ between CMB data and DESI, can lead to significant shifts in $n_s$ when the two are combined.

Determinations of $n_s$ from CMB data are in agreement: the ACT primary CMB constraint  $n_s = 0.9667 \pm 0.0077$ can be compared with the {\it Planck} 2018 PR3 ({\texttt{plik}}) constraint $0.9649\pm 0.0044$ and the SPT-3G primary CMB constraint $n_s=0.948\pm 0.012$. The latest SPT data release  \cite{SPT-3G:2025bzu} found that the combination of data from {\it Planck}, ACT, and SPT gives $n_s = 0.9684\pm 0.003$, which is compatible with the inflationary models mentioned above. 
Adding BAO data, these constraints generally shift upwards as mentioned earlier in this Introduction.

We emphasize that the BAO data place no constraint on $n_s$ directly. Their contribution to the $n_s$-constraining power of the combination of CMB and BAO data derives solely from correlations between uncertainties in the CMB-only inferences of $n_s$ with uncertainties in the CMB-only inferences of $r_dh$ and $\Omega_m$.

To contextualize the $n_s$ constraints presented in this paper, we will draw frequent comparison to the `benchmark models' of inflation considered by CMB-S4 and LiteBIRD B-mode searches, namely the Starobinsky model, Higgs inflation, and $\alpha$-attractors, all of which, in the large $N_*$ approximation, predict $n_s$ given by \cite{Kallosh:2025ijd}
\begin{equation}\label{largeN}
    n_s = 1 - \frac{2}{N_*} \ .
\end{equation}
Here $N_*$ denotes the number of efolds before the end of inflation that the CMB pivot scale, 
conventionally taken to be $k$ = 0.05/Mpc,
exited the horizon. The predicted range of $N_*$ (and small deviations from the asymptotic relation (\ref{largeN})) depends on the inflation model and the reheating model. In this work, we take the model-agnostic range $N_*=[50,60]$, corresponding to a range in $n_s$ of $[0.9600,0.9667]$.

The structure of this paper is as follows: In Sec.~\ref{sec:datasets} we introduce the datasets and likelihoods to be used in our analyses. In Sec.~\ref{sec:BAOns} we perform a detailed study of the interplay between BAO data and CMB data and the consequent $n_s$ constraints when BAO and CMB datasets are combined. In Sec.~\ref{subsec:BAOtension} we review the tension between CMB experiments and DESI DR2 and highlight the correlations between BAO parameters and $n_s$. In Sec.~\ref{subsec:ACTDESI} we perform the first MCMC analysis of the joint dataset of ACT primary CMB combined with DESI DR2 (``ACT+DESI'') and compare the $n_s$ constraint to that from ACT alone, with results consistent with that anticipated from the BAO-CMB tension in Sec.~\ref{subsec:BAOtension}. In Sec.~\ref{subsec:OtherCMB} we compare other CMB datasets, and in particular why the ACT $n_s$ correlations with BAO parameters are weaker than the Planck $n_s$ correlations with BAO parameters. Finally, in Sec.~\ref{sec:inflation} we consider the implications of these 
various constraints for inflationary models. We close in Sec.~\ref{sec:discussion} with a discussion of directions for future work.

\section{Data sets and methodology}
\label{sec:datasets}

We perform Markov Chain Monte Carlo analyses of CMB and BAO data to obtain constraints on cosmological parameters using \texttt{Cobaya} \cite{torrado_lewis_2019}\footnote{\url{https://github.com/CobayaSampler/cobaya/tree/master}}. We run the chains until the Gelman-Rubin convergence parameter, $ R-1 <0.01$. We compute theoretical predictions using the Boltzmann solvers \texttt{CAMB} ~\cite{Lewis:1999bs} and \texttt{CLASS} ~\cite{blas11}.\footnote{We use the same accuracy settings as~\cite{ACT:2025fju, SPT-3G:2025bzu}} We use \texttt{GetDist} ~\cite{GetDist}\footnote{\url{https://github.com/cmbant/getdist}} to analyze and plot the MCMC runs. We use the \texttt{BOBYQA} minimizer~\cite{Cartis:2018jxl,Cartis:2018xum}, implemented in \texttt{Cobaya}, to compute the best-fit parameters initialized from maximum a posteriori (MAP) points of the corresponding MCMC chains.

We use the latest BAO likelihood supplied by the DESI collaboration based on \cite{DESIDR2}.
We use the following CMB data sets:
\begin{itemize}
    \item ACT DR6 primary CMB (MFLike), \footnote{\url{https://act.princeton.edu/act-dr6-data-products}} \cite{ACT:2025fju}.  In what follows, when we refer to ACT DR6 we do not include the ACT CMB lensing reconstruction unless otherwise noted.
    \item \textit{Planck}, specifically the 2018 primary CMB and lensing data based on the PR3 \cite{planck18-5}. 
\end{itemize}
We focus our attention on ACT in light of the significant attention ACT has received from the inflation community, e.g.~Refs.~\cite{Kallosh:2025rni,Maity:2025czp,Heidarian:2025drk,Frolovsky:2025iao,Chakraborty:2025jof,Lizarraga:2025aiw,Cacciapaglia:2025xqd,Addazi:2025qra,Kim:2025dyi,Pallis:2025epn,Odintsov:2025wai,Gialamas:2025kef,McDonald:2025odl,Yin:2025rrs,Aoki:2025wld,Antoniadis:2025pfa,Haque:2025uis,He:2025bli,Salvio:2025izr,Pallis:2025nrv,Haque:2025uga,Hai:2025wvs,Saini:2025jlc,Kohri:2025lau,Liu:2025qca,Drees:2025ngb,Zharov:2025evb,Yogesh:2025wak,Byrnes:2025kit,Okada:2025lpl,Byrnes:2025kit,Yogesh:2025wak,Peng:2025bws,McDonald:2025tfp,Choudhury:2025vso,Mohammadi:2025gbu}, and for a timely release of this paper, though we note that new and relevant SPT results \cite{SPT-3G:2025bzu} came out quite recently, and a deeper comparison with SPT would be warranted in a future publication.
We compare 
{\it Planck} and ACT in Sec.~\ref{subsec:OtherCMB}.

When performing MCMC analyses of ACT, we follow the ACT convention and include the {\it Planck}-based low-$\ell$ EE power spectrum Sroll2 likelihood \cite{ACT:2025tim}\footnote{If we adopt a tau prior instead of using {\it Planck}-based low-$\ell$ EE power spectrum Sroll2 in the ACT + DESI DR2 runs, the result is unchanged. }.
When possible, we use publicly available MCMC runs for ACT and {\it Planck} data.
Though the optical depth to reionization is relevant in the comparison of CMB and BAO data, the different choices above lead to small changes that do not change the key message of this work (see \S VII C and Appendix I in \cite{SPT-3G:2025bzu} for details).

\section{The BAO-CMB Tension and Impact On $n_s$}
\label{sec:BAOns}

\subsection{The BAO-CMB Tension}
\label{subsec:BAOtension}

As introduced above, in what follows we refer to the {\it BAO-CMB tension} as that between the CMB and DESI BAO inferences of the parameters $r_d h$ and $\Omega_m$. This tension was highlighted in \cite{SPT-3G:2025bzu}, where it was shown to range from $\approx 2-4\sigma$ depending on the CMB dataset(s) considered. In the case of ACT, the tension is $3.1\sigma$, which increases to $3.7\sigma$ if ACT is combined with SPT.
\footnote{Note that certain analysis choices can influence this comparison.
Specifically, the choice of external constraint on the optical depth to reionization or addition of certain lensing data can reduce the BAO-CMB tension for ACT to the $2.7\,\sigma$ level\cite{SPT-3G:2025bzu, ACT:2025fju, desi_act}. Though formally below the $3\,\sigma$ threshold used by \cite{SPT-3G:2025bzu}, the differences between ACT and DESI remain substantial.}

To understand the focus on the parameters $r_dh$ and $\Omega_m$, it is helpful to return to first principles.
Analogous to the role of last scattering in the evolution of CMB photons, the BAO are frozen in at the end of the baryon drag epoch, when baryons decouple. They imprint a comoving length scale in galaxy 2-point correlation functions equal to the comoving sound horizon at that time,  $r_d$. This scale is observable in redshift space as $\Delta z_s(z) = r_d H(z)$ along the line of sight and $\Delta \theta_s(z) = r_d/D_M(z)$ transverse to the line of sight, where $D_M(z)$ is the comoving angular-diameter distance. It is straightforward to show that, in the $\Lambda$CDM model, $\Delta z_s(z)$ and $\Delta \theta_s(z)$, as inferred from BAO surveys, are determined to high accuracy by specifying just two parameters that one can take to be $\Omega_{m}$ and $r_d h$ where $h = H_0/(100 {\rm km/sec/Mpc})$. It follows that, in the context of $\Lambda$CDM, BAO data can be losslessly compressed into constraints on $r_dh$ and $\Omega_m$.

\begin{figure*}
    \centering
    \includegraphics[scale=0.5]{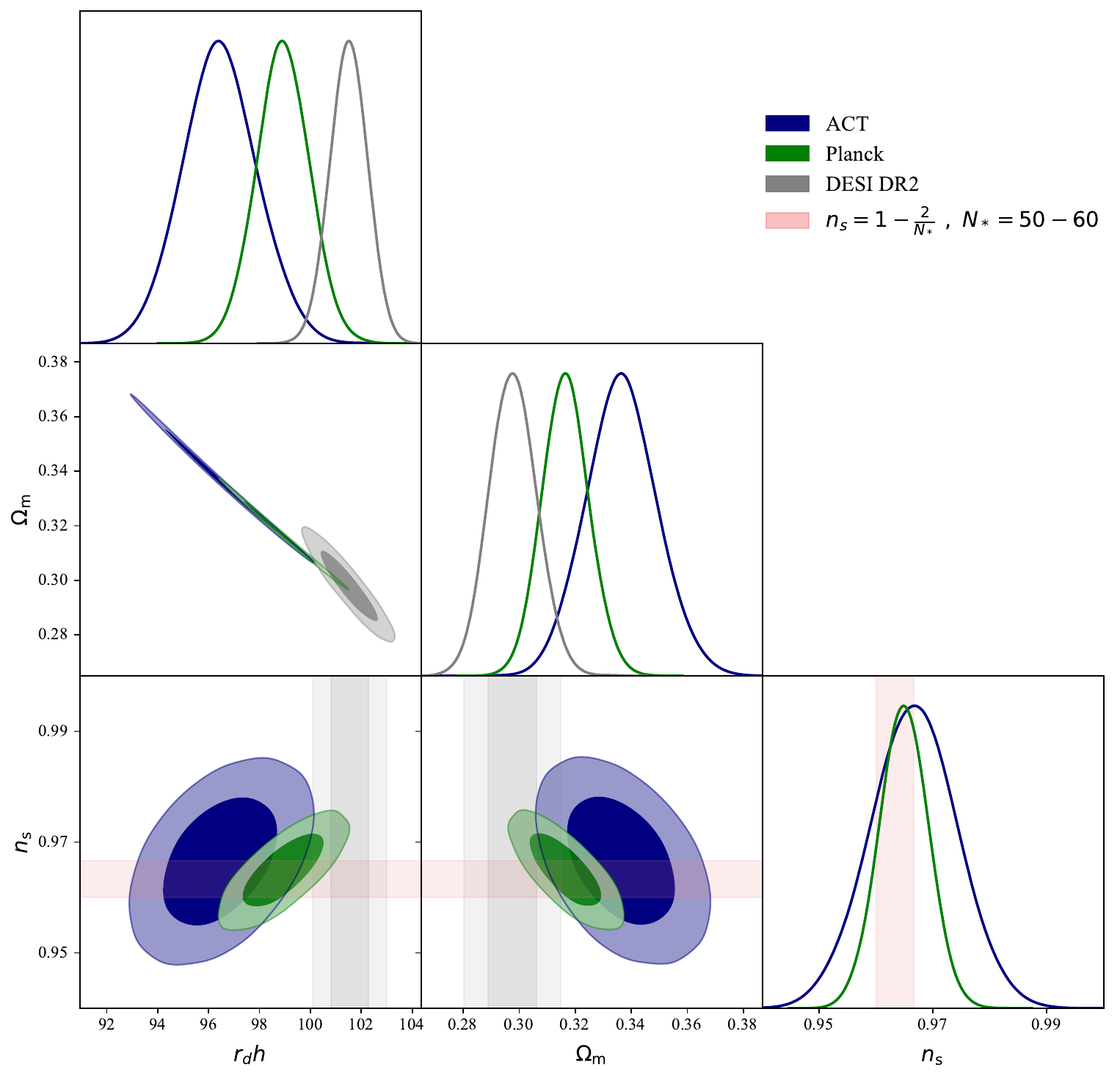}
    \caption{Comparison of CMB data sets: Constraints from ACT and {\it Planck}. Overlaid are constraints on BAO parameters from DESI DR2 (grey bands). Pink shaded bands show the spectral index $n_s=1-2/N_*$, as predicted by the Starobinsky, Higgs, and exponential $\alpha$-attractor models, for $N_*=50-60$. }
    \label{fig:ACT-Planck-SPT}
\end{figure*}

The BAO parameters $r_dh$ and $\Omega_m$ can also be inferred from CMB data, leading to correlations with other $\Lambda$CDM parameters, including $n_s$. The matter density $\omega_m \equiv \Omega_m h^2$ is determined primarily by the imprint of the `radiation driving' effect, a boosting of acoustic oscillation amplitude due to potential decay at horizon crossing \citep{Hu:1996mn,Planck:2016tof}. The amplitude depends on the matter-to-radiation ratio at horizon crossing and hence changes smoothly over a wide range of angular scales, with some similarity to changes in $n_s$, leading to a partial degeneracy between $n_s$ and $\omega_m$. Determination of $\Omega_m$ comes from combining $\omega_m$ and the angular size of the sound horizon, $\theta_s$. The latter is determined with extremely small uncertainty due to the fact that it determines the spacing between acoustic peaks. In the $\Lambda$CDM model a constraint on $\theta_s$ leads to a constraint on $\Omega_m h^3$ \citep{2dFGRSTeam:2002tzq}. The combination thus can deliver $\Omega_m = (\Omega_m h^2)^3 / (\Omega_m h^3)^2.$ This $\Omega_m$ determination is correlated with $n_s$ since $\Omega_m h^2$ is correlated with $n_s$ and $\theta_s$ is not.

In contrast, both $h$ and $r_d$, and therefore $r_d h$, are positively correlated with $n_s$. We expect these correlations due to the correlation between $n_s$ and $\omega_m$ just described and the dependence of $h$ and $r_d$ on $\omega_m$.  Since $h = \Omega_m h^3/\omega_m$ we expect $\delta h/h = -\delta \omega_m/\omega_m$ and in \cite{Hu:2000ti} it was shown that $\delta r_d/r_d = -1/4\,  \delta \omega_m/\omega_m$.\footnote{The -1/4 would be -1/2 in the absence of the radiation contribution to the Friedmann equation, which softens the response. This scaling was derived for $r_s$, the comoving size of the sound horizon at last scattering (rather than the end of the baryon drag epoch), but the change in redshift is small, and we expect that the scalings are very similar.}

We begin our analysis of the BAO-CMB tension and the interplay with $n_s$ with an examination of CMB constraints in comparison to those from DESI DR2. In Fig.~\ref{fig:ACT-Planck-SPT} we show the marginalized posterior distributions for the BAO parameters and the spectral index $n_s$, in the fit of the $\Lambda$CDM model to primary CMB data from ACT DR6 and {\it Planck} 2018. The DESI DR2 constraints on $r_dh$ and $\Omega_m$ are shown in grey and green bands, respectively, and the prediction of the benchmark inflationary models (Eq.~\eqref{largeN}) is shown in pink bands. From this, one may appreciate the correlation between $\Omega_m$ and $r_dh$, and the correlation of each of these with $n_s$: moving $r_dh$ and $\Omega_m$ towards the DESI-preferred region, while maintaining the fit to the data, requires a commensurate increase in $n_s$.

\begin{figure}[h!]
    \centering
    \includegraphics[width=0.5\textwidth]{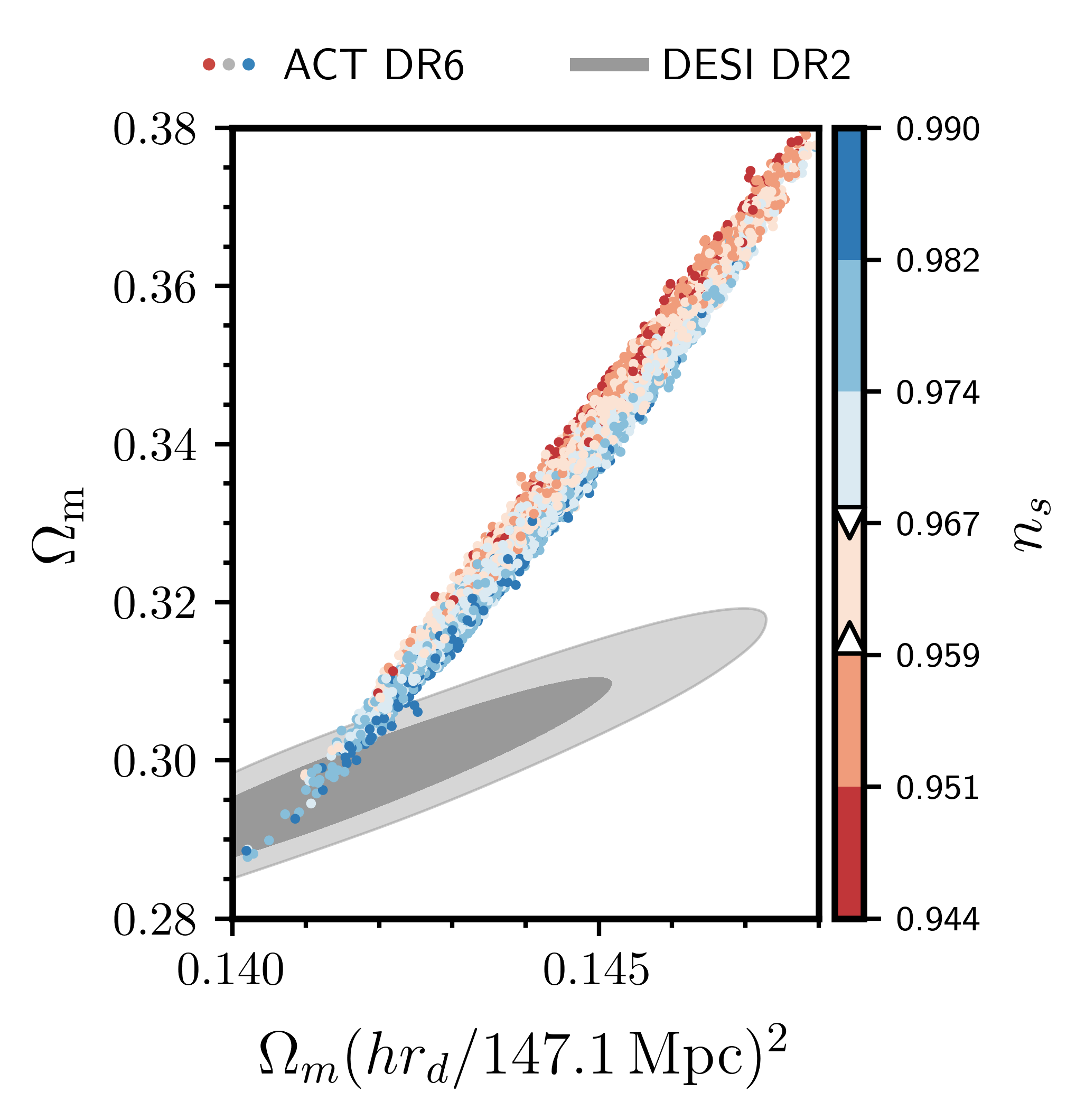}
    \caption{Visualizing $n_s$ in the $\Omega_m$-$\Omega_m (r_d h/147.1\, \mathrm{Mpc})^2$ plane. We plot the publicly available ACT primary CMB constraint on  $\Omega_m$-$\Omega_m (r_d h/147.1\, \mathrm{Mpc})^2$  with MCMC samples color-coded by their $n_s$ values: color map centered on the mean of the posterior and incrementing in $1\,\sigma$ intervals. The lightest red shade matches almost exactly the range of $n_s$ values predicted by benchmark inflationary models: we indicate $n_s=1-2/N_\star$ for $N_\star=50$ and $60$ with an upward and downward-pointing triangle, respectively. DESI DR2 constraint is in grey. ACT barely enters DESI, and when it does, $n_s \gtrsim 0.97$.}
    \label{fig:ns-color-coded}
\end{figure}

To dig deeper into the correlations with $n_s$, in Fig.~\ref{fig:ns-color-coded} we show the $\Omega_m-r_dh$ constraint from ACT and from DESI DR2, where $n_s$ values of samples in the publicly available MCMC chains are shown in colored dots. To add clarity in light of the tight correlation, we show constraints in the plane of $\Omega_m$-$\Omega_m (r_d h/147.1\, \mathrm{Mpc})^2$. From this, one can see that the only values of $n_s$ within the DESI region are much higher than the central value.

We note that the marginalized $n_s$ constraint is in good agreement with the Starobinsky model, Higgs inflation, and $\alpha$-attractors for both ACT and {\it Planck}.

However, we emphasize that in the $n_s-r_dh$ plane, there is no simultaneous overlap between the ACT 2$\sigma$ contour, the 2$\sigma$ DESI constraint on $r_dh$, and the band of $n_s$ values predicted by benchmark inflation scenarios.
This incompatibility suggests that there is little evidence that ACT and DESI constraints are both right, while at the same time $\Lambda\mathrm{CDM}$ with a benchmark inflation scenario holds true; something has to give.

Inspecting the correlations in the parameter space (see again Fig.~\ref{fig:ns-color-coded}), where ACT data are most compatible with DESI, it is far from the benchmark inflationary models, and where ACT data are compatible with benchmark inflationary models, it is far from matching DESI. 
From this, one may anticipate that the combination of ACT with DESI, given the $\approx 3\sigma$ tension between the two data sets, will drive $n_s$ away from the benchmark inflation region.


\subsection{Shift in $n_s$ from DESI BAO data}

\label{subsec:ACTDESI}

To test the intuition built in the previous section for $n_s$ when CMB and DESI are combined, we perform an analysis of ACT DR6 (primary CMB) and DESI DR2 BAO data together. The use of ACT data as a case study is motivated by the significant attention ACT has received from the inflation community \cite{Kallosh:2025rni,Maity:2025czp,Heidarian:2025drk,Frolovsky:2025iao,Chakraborty:2025jof,Lizarraga:2025aiw,Cacciapaglia:2025xqd,Addazi:2025qra,Kim:2025dyi,Pallis:2025epn,Odintsov:2025wai,Gialamas:2025kef,McDonald:2025odl,Yin:2025rrs,Aoki:2025wld,Antoniadis:2025pfa,Haque:2025uis,He:2025bli,Salvio:2025izr,Pallis:2025nrv,Haque:2025uga,Hai:2025wvs,Saini:2025jlc,Kohri:2025lau,Liu:2025qca,Drees:2025ngb,Zharov:2025evb,Yogesh:2025wak,Byrnes:2025kit,Okada:2025lpl,Byrnes:2025kit,Yogesh:2025wak,Peng:2025bws,McDonald:2025tfp,Choudhury:2025vso}. We compare to other CMB data sets in Sec.~\ref{subsec:OtherCMB}.

\begin{figure*}
    \centering
    \includegraphics[scale=0.5]{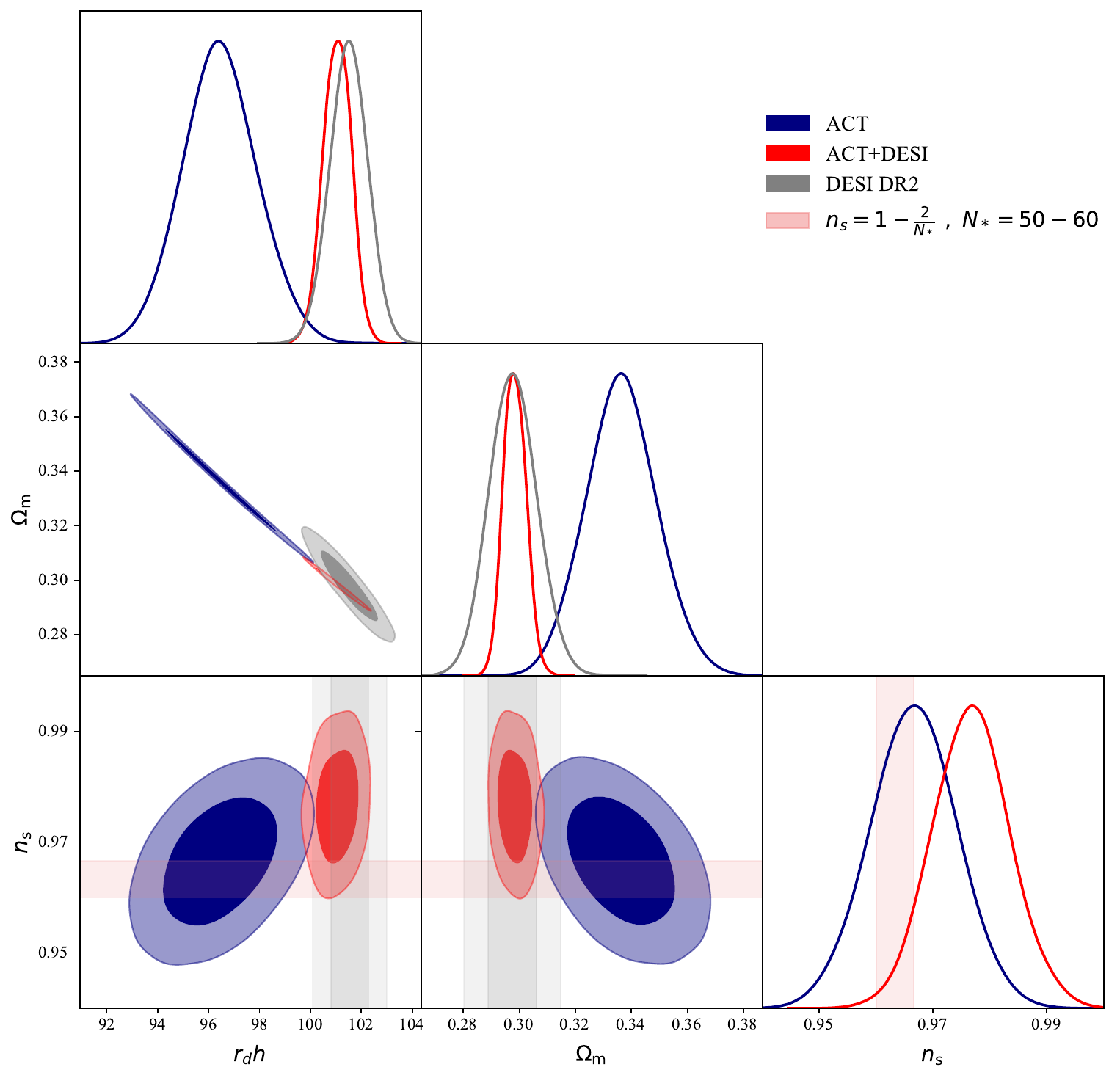}
    \caption{Same parameter planes and plot elements for ACT, DESI, and $n_s$ benchmark values as Figure~\ref{fig:ACT-Planck-SPT}. We now add the joint ACT+DESI results in red. The correlations between $n_s$ and BAO observables for ACT provide a lever arm for the BAO-CMB tension and lead to an upward shift in $n_s$ in the joint constraints. These correlations are a feature common to CMB data sets that are weighted towards low to mid $\ell$ (such as \textit{Planck}) and are not unique to the ACT data.}
    \label{fig:ACT-DESI}
\end{figure*}

The analysis presented here differs from the baseline analyses performed in \cite{ACT:2025fju} where DESI (DR1) is included only in combination with ACT, {\it Planck} ($\ell < 1000$), and CMB lensing.
\cite{SPT-3G:2025bzu} do not report results for the combination of ACT DR6  and DESI DR2 due to the differences between the two data sets\footnote{though the comparison made in \cite{SPT-3G:2025bzu}  is between DESI and ACT DR6 {\it including} CMB lensing}; furthermore, this joint analysis is also not performed in the updated appendix of \cite{ACT:2025fju}.
Here, we intentionally forego the cautious approach taken by these works in order to better understand shifts to the scalar spectral index that arise in joint analyses of ACT and DESI data and to caution against the interpretation of these results in the context of inflation.
Results are shown in Fig.~\ref{fig:ACT-DESI}, parameter constraints are given in Tab.~\ref{table:constraints-ACT-DESI}, and the $\chi^2$ statistics are tabulated in Tab.~\ref{table:chi2_ACT_DESI}.

Comparing the posterior distributions in the fit to ACT with those of ACT combined with DESI, shown in Fig.~\ref{fig:ACT-DESI}, one may appreciate a significant shift in the BAO parameters $r_dh$ and $\Omega_m$ when DESI is included, with a commensurate shift of $n_s$ along the degeneracy direction of the fit to ACT alone. The constraint on $n_s$ changes from $n_s=0.9666 \pm 0.0076$ from ACT alone to $n_s=0.9770  \pm 0.0070$ when ACT is combined with DESI DR2.

\begin{table}[htb!]
Constraints from ACT DR6 CMB and DESI DR2 BAO Data \\
  \centering
  \begin{tabular}{|l|c|c|c|}
    \hline\hline Parameter &ACT  ~~& ~~~ACT + DESI \\ \hline \hline

    $n_\mathrm{s}$ & $0.9666 \, ( 0.9664 )\pm 0.0076$ &$0.9770 \, (0.9754) \, \pm 0.0070$   \\

   $r_d h$ [Mpc] & $96.5 \, (96.27) \,\pm 1.5$ &  $101.04 \, (101.05) \pm 0.54 $ \\
   $\Omega_m$ & $0.337 \, ( 0.338) \, \pm 0.013$ & $0.2999 \, (0.2996) \, \pm 0.0040  $\\
   \hline
    \hline
  \end{tabular} 
  \caption{Parameter constraints (best fit) from the fit of $\Lambda$CDM to ACT DR6 primary CMB data alone and in combination with DESI BAO data.}
  \label{table:constraints-ACT-DESI}
\end{table}

\begin{table}[h!]
\centering
$\chi^2$ statistics from ACT and DESI
  \begin{tabular}{|l|c|c|}
    \hline\hline
    Datasets & ACT & ACT+DESI\\ \hline \hline
        ACT DR6  & 5881.3  & 5888.8\\
        {\it Planck} low-$\ell$ EE (Sroll2) & 390.0  & 391.0 \\
        DESI DR2 & -- & 11.5\\
    \hline
    $\Delta \chi^2 _{\rm CMB} $ &  & +8.5\\ 
    \hline
  \end{tabular}
  \caption{ $\chi^2$ values for the best-fit $\Lambda$CDM model in the fit to the ACT primary CMB alone and in combination with DESI DR2. 
  Although the CMB $\chi^2$ increases by $8.5$ points, the number of degrees of freedom of the underlying distribution also changes; an interpretation of the $\Delta \chi^2$ in terms of the fit quality is therefore not straightforward. 
   Note ``ACT" always refers to ACT + {\it Planck} low-$\ell$ EE following the ACT convention \cite{ACT:2025tim}.}
  \label{table:chi2_ACT_DESI}
\end{table}

\subsection{Comparison of CMB data sets}
\label{subsec:OtherCMB}

To conclude our analysis, we compare ACT to other CMB data sets.
As others have pointed out, ACT data are in excellent agreement with SPT and {\it Planck} in $\Lambda\mathrm{CDM}$ and in extended model spaces \cite{SPT-3G:2025bzu, ACT:2025tim, ACT:2025fju}.

Further comparison is facilitated by Fig.~\ref{fig:ACT-Planck-SPT}. Note that the central value of the {\it Planck} $r_dh$ and $\Omega_m$ contour is closer to the DESI-preferred region than are the corresponding ACT constraints, consistent with a smaller BAO-CMB tension in the case of {\it Planck}, as reported by \cite{SPT-3G:2025bzu}. 
Moreover, we note the differing degree of correlation of BAO parameters with $n_s$. The correlation structure depends non-trivially on the angular scales covered by the power spectrum measurements, the weighting of the data towards temperature or polarization set by the noise level, as well as on the relative weight of gravitational lensing information. Note that the experiment with more of its information coming from large angular scales ({\it Planck}) has the higher correlation. 
This is not surprising for two reasons. First, for experiments without sensitivity to larger angular scales, the `radiation driving' effects, discussed briefly in Sec.~\ref{sec:BAOns}, are less informative about $\Omega_m h^2$ since the envelope asymptotes to a fixed constant value on small scales, obscuring the necessary envelope shape information. Second, gravitational lensing is more important on smaller scales, and is another source of information about $\Omega_m h^2$, one we expect to be relatively uncorrelated with $n_s$.

\section{ACT, DESI, and inflationary models}
\label{sec:inflation}

 We now turn to the implications of these various $n_s$ constraints for models of inflation. To do so it is important to first recall the results of the {\it Planck} experiment.

\begin{figure}[h]
\centering
\includegraphics[scale=0.48]{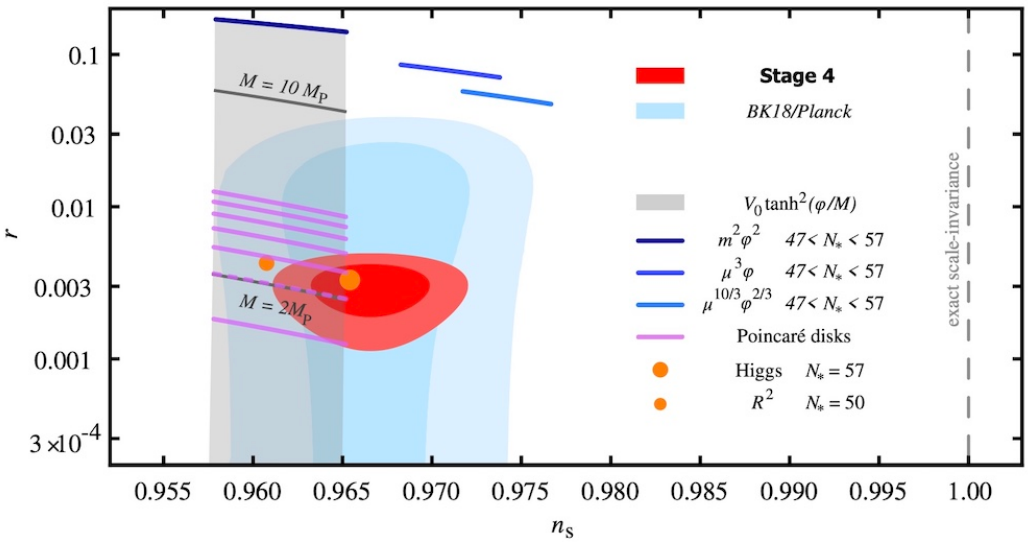}
\caption{\footnotesize The figure from the CMB White Paper \cite{Chang:2022tzj}. It shows the predictions of  $\alpha$-attractors $\tanh^2 {\vp/M}$ with $M = \sqrt{6\alpha}$ (gray band), the predictions of $\alpha$-attractors for Poincar\'e disks with integer  $3\alpha$ (purple lines), as well as Higgs inflation and the Starobinsky ($R^2$) model (orange circles).  Starobinsky and Higgs inflation correspond to $\alpha$-attractors with $\a=1$. The predictions are for $47 < N_*< 57$. If one plots the model predictions in the often used range $50 < N_*< 60$, they would shift to the right by about $0.002$. Note a complete absence of viable targets at $n_s \gtrsim 0.967$.} 
 \label{Flauger0}
\end{figure}

In 2013-2015, {\it Planck} sent a simple message: the data suggest that concave potentials are preferred over convex potentials \cite{Planck:2013jfk,Planck:2015sxf}.
 In particular, it implied that {\it plateau potentials are preferable to monomial potentials. Moreover, over the last 10 years, the data have supported models approaching the plateau exponentially.} This is why in the CMB White Paper \cite{Chang:2022tzj} and LiteBIRD \cite{LiteBIRD:2022cnt} figures we see only 3 viable targets: the Starobinsky model \cite{Starobinsky:1980te}, Higgs inflation \cite{Salopek:1988qh,Bezrukov:2007ep} and  $\alpha$-attractors \cite{Kallosh:2013yoa}; see for example Fig.  \ref{Flauger0} here, which is Fig. 2  in \cite{Chang:2022tzj}. 
All of these models have potentials that exponentially approach a plateau at large values of the inflaton field $\vp$, so one may call them exponential attractors. These models include T and E $\alpha$-attractor models  \cite{Kallosh:2013yoa} with potentials 
   \be
V_T(\phi) = V_{0} \tanh^{2n}{\varphi\over\sqrt {6 \alpha}}\, , \qquad V_E = V_0 \Bigl(1 - e^{-\sqrt {2 \over 3\alpha}\varphi} \Bigr)^{2n} \ .
\label{exp}\ee 
In both cases, the potentials at large $\vp$ can be represented as
\be\label{plateau1}
V(\vp) = V_{0}(1 - C \ e^{-\sqrt{2\over 3\alpha} \varphi }+\dots ) \ .
\ee
The constant $C$ can be absorbed into a redefinition (shift) of the field $\vp$.  All inflationary predictions in the regime with $e^{-\sqrt{2\over 3\alpha} \varphi } \ll 1$ do not depend on $n$, and are determined by two parameters, $V_{0}$ and $\alpha$:
\be
\label{pred}
 A_{s} \approx {V_{0}\, N_*^{2}\over 18 \pi^{2 }\alpha} \ , \qquad n_{s} \approx 1-{2\over N_*} \ , \qquad r \approx {12\alpha\over N^{2}_{*}} \ .
\ee 
Accuracy of these results increases with an increase of $N_*$ and a decrease of $\alpha$, i.e., at small $r$. These results are compatible with all currently available {\it Planck}/BICEP/Keck/ACT/SPT data, prior to the addition of DESI data. 

The potential of the E-model with $n = 1$, $\alpha=1$ coincides with the potential of the Starobinsky model. The Higgs inflation potential belongs to the class of T-models with $\alpha = 1$. The models with $\alpha=2$ can be associated with Fibre inflation in string theory, see \cite{Cicoli:2024bxw} and references therein. Discrete values of $3\a=1, \dots , 7$ are Poincar\'e disk targets \cite{Ferrara:2016fwe}.   All of these models typically predict $0.96< n_{s} < 0.967$, but there are some model-dependent variations.

For example, simplest reheating mechanisms in the Starobinsky model are not very efficient, therefore one typically has $n_{s} \sim 0.962 - 0.963$, which is reflected in Fig. \ref{Flauger0}. In the  Higgs inflation model, reheating is very efficient because the Higgs field is a part of the Standard Model, which is why Fig.  \ref{Flauger0} shows it at $n_{s}\sim 0.966$. Reheating in $\alpha$-attractors is somewhat less constrained, depending on the specific model version. In particular, E-models introduced in  \cite{Kallosh:2013yoa} can reach $n_{s} \approx 0.97$ for $N_* = 60$. There is also a class of quintessential $\alpha$-attractors, where the potential may not have a minimum. Such models may describe either a cosmological constant or dark energy, and they may have $n_{s}$ higher than that in more conventional models by $\Delta n_{s} \sim 0.006$ \cite{Akrami:2017cir}. 

If one takes the P-ACT-LB result $n_s=0.9739\pm 0.0034$ 
at
face value, it removes all viable targets shown in the CMB-S4 and LiteBIRD figures (which do not include E-models and quintessential attractors), and forces us to reconsider the situation and explore the models that could describe inflation with $n_{s} \gtrsim 0.97$.

One should note that the situation is less dramatic than it could seem because we have simple inflationary models with polynomial potentials which can account for {\it any} values of the 3 main CMB-related parameters $A_{s}$, $n_{s}$ and $r$ by a choice of 3 parameters in the models. For example, consider the potential  $V = {m^{2}\phi^{2}\over 2 } (1-a\phi + b (a\phi)^{2})^{2} $, which naturally appears in the simplest versions of chaotic inflation in supergravity \cite{Kallosh:2014xwa}. For $N_* = 60$, $b = 0.34$, and $a = 0.13$ the predictions nearly coincide with the predictions of the Starobinsky model, the Higgs inflation model, and  $\alpha$-attractors with $\alpha = 1$: \ $n_{s}= 0.967$, $r \approx 3\times 10^{{-3}}$.  For $a = 0.14$, the value of $n_{s}$ becomes $0.973$, matching the 
P-ACT-LB
result. Further increase of $a$ practically does not change $r$, but gradually moves $n_{s}$ all the way to $n_{s} = 1$. By a slight change to $b$, one can dial to any value of $r$, and then get a desirable value of $n_{s}$ by changing $a$ \cite{Kallosh:2025ijd}. 

Then what is the problem? First of all, by enabling us to describe any set of parameters $A_{s}$, $n_{s}$ and $r$ (which is not a small feat!), this model does not provide any particular guidance to observers and does not relate the observations to fundamental physics.   Meanwhile,  exponential $\alpha$-attractors describe $n_{s}$ and $r$ by a choice of a single parameter $\a$, with $\a$-independent $n_s$ and $r$ proportional to $\a$. 

In all $\a$-attractor models, the parameter $\a$ has geometric origin: the scalar kinetic term is defined by hyperbolic geometry, an intrinsic property of extended supergravity.  $3\a$ is related to the  \K curvature $\mathcal{R}_K$ of the underlying Poincar\'e disk geometry,  
$ds^2= 3\a {dx^2+dy^2\over (1- x^2-y^2)^2}$ \cite{Ferrara:2013rsa},
$3\a= |2/\mathcal{R}_K|$, and  $r= {8\over |\mathcal{R}_K| N_*^2}$. 

\begin{figure}[h]
\centering
\includegraphics[scale=0.153]{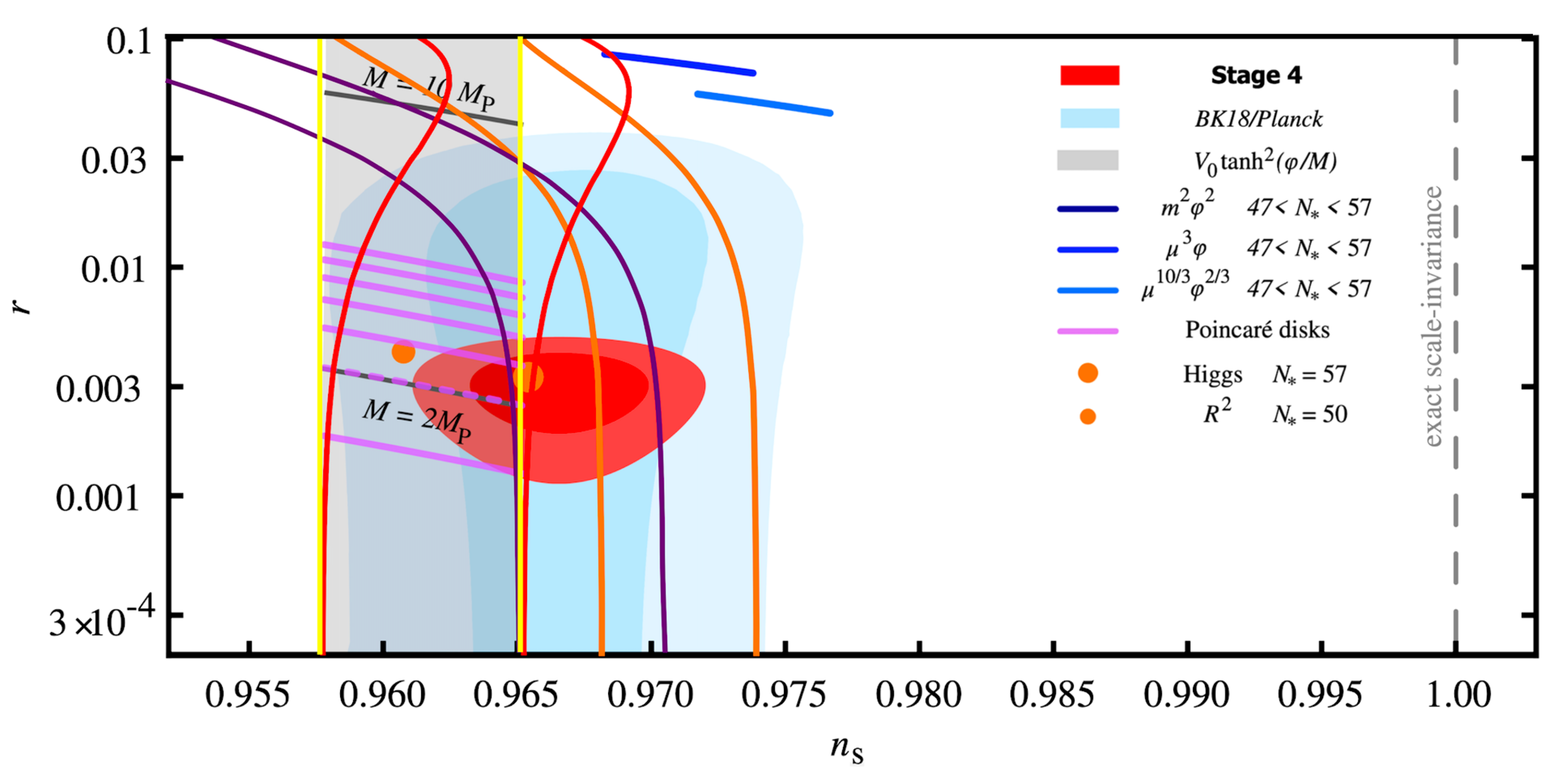}
\caption{\footnotesize An extended version of Fig. \ref{Flauger0}, which is  Fig. 1 in   \cite{Kallosh:2022vha}.   In addition to the targets in Fig. \ref{Flauger0} here, the red lines represent the E-model $\a$-attractors, while the purple and orange lines show the predictions of the polynomial attractors with $k=4$ and $k=2$, respectively.  Polynomial $\alpha$-attractors have a power law approach to plateau, $V=V_0 (1- \big(\mu/ \vp\big)^k+\dots)$. Predictions of the simplest models with $k=4$ and $k=2$ in the small $r$ area have $0.965< n_{s}< 0.974$ for $47 < N_*< 57$.} 
\label{Flauger}
\end{figure}

Another set of models with some attractive features, but predicting $n_s \gtrsim  0.97$,  has been studied over the years. They includes pole inflation \cite{Galante:2014ifa,Kallosh:2019hzo}, D-brane inflation \cite{Dvali:1998pa,Burgess:2001fx}, KKLTI models   \cite{Kachru:2003sx,Martin:2013tda} and polynomial $\alpha$-attractors \cite{Kallosh:2022feu},  see the recent review \cite{Kallosh:2025ijd} for a more detailed discussion. All of these models have one common feature: their potentials follow a power law approach to the plateau, $V=V_0 (1- \big(\mu/ \vp\big)^k+\dots)$. Predictions of these models with $k=4,2$ are shown in Fig. \ref{Flauger}.

Here are some potentials which appear in the polynomial $\alpha$-attractors:  
 \be
V(\varphi) =V_0 {|\varphi|^k\over \mu^k + |\varphi|^k}, \qquad V\to V_0 (1- ( \mu / |\vp| )^k+\dots)  \, , 
\label{polynomial}\ee 
where $\mu = \sqrt{3\alpha\,\over 2}$. At small $\a$ and large $N_*$ in the attractor regime one finds that \cite{Martin:2013tda,Kallosh:2018zsi}
\be
n_{s} \approx 1-{\beta \over N_*}, \quad r\approx 8k^2 {\mu^{2k\over k+2}\over (k(k+2) N_*)^\beta}, \quad \beta ={2(k+1)\over k+2} \ .
\label{nsr}\ee 
In polynomial $\alpha$-attractors  there is an  $\a$-independent $n_s$ and  $\a$-dependent   parameter $r$, and both parameters depend on $k$, the order of a polynomial approach to the plateau.

For $k=4,3,2,1$ we have $\beta = 5/3, 8/5, 3/2, 4/3$. For all $k$ one has  $\beta<2$   \cite{Kallosh:2019hzo}, so that decreasing $k$ moves the $n_s$ to the right.
Meanwhile, $r$ decreases with the decrease of $\a$, as we see from the equations  \rf{polynomial}, \rf{nsr}.  For the polynomial  $\a$-attractor models, just as for the exponential attractors, there is a set of discrete targets for $r$ corresponding to the Poncar\'e disks with $3\a=1,2,3,4,5,6,7$.
In polynomial $\a$-atractors $\mu^2= {3\a\over 2} ={1\over  |\mathcal{R}_K|}$.

It is fair to say that all of these models have some disadvantages compared to the Starobinsky model, Higgs inflation, and exponential $\alpha$-attractors.  For example, models of D-brane inflation based on string theory require very small values of $\mu$ corresponding to $r\lesssim 10^{-5}$, which cannot be a realistic target for B-mode searches. Out of all poll inflationary models, only $\alpha$-attractors have an interesting geometric interpretation. Polynomial $\alpha$-attractors require potentials with some specific properties discussed in \cite{Kallosh:2022feu}, which makes them less general. But they have an advantage: just as models of exponential $\alpha$-attractors, they can be embedded into extended supergravity. Polynomial models also predict the values of $r$ decreasing with decreasing $\a$ and increasing \K curvature.

Thus,  in all $\a$-attractors, exponential and polynomial, the tensor-to-scalar ratio $r$  decreases when the moduli space curvature $ {\cal R}_K $ increases. Therefore, the future detection of B-modes in the context of $\a$-attractors may provide some information about the moduli space curvature in associated cosmological inflationary models.  In this way, one can relate discoveries in observational cosmology to fundamental physics.

 If instead of $47 < N_*< 57$ one considers the range $50 < N_*< 60$, as in  {\it Planck},  ACT and SPT data releases,  the boundary of the gray area in  Fig. \ref{Flauger} moves to $n_{s} = 0.967$, and the right orange line of $k=2$ model in Fig. \ref{Flauger} moves to an attractor value of $n_s = 0.975$. By considering non-integer values of $k$,  for $N_* = 60$ one can cover all values of $n_{s}$ from $1-2/N_*\sim 0.967$   to $1-1/N_*  \sim 0.983$.
 
\begin{figure}[h]
\centering
\includegraphics[scale=0.3]{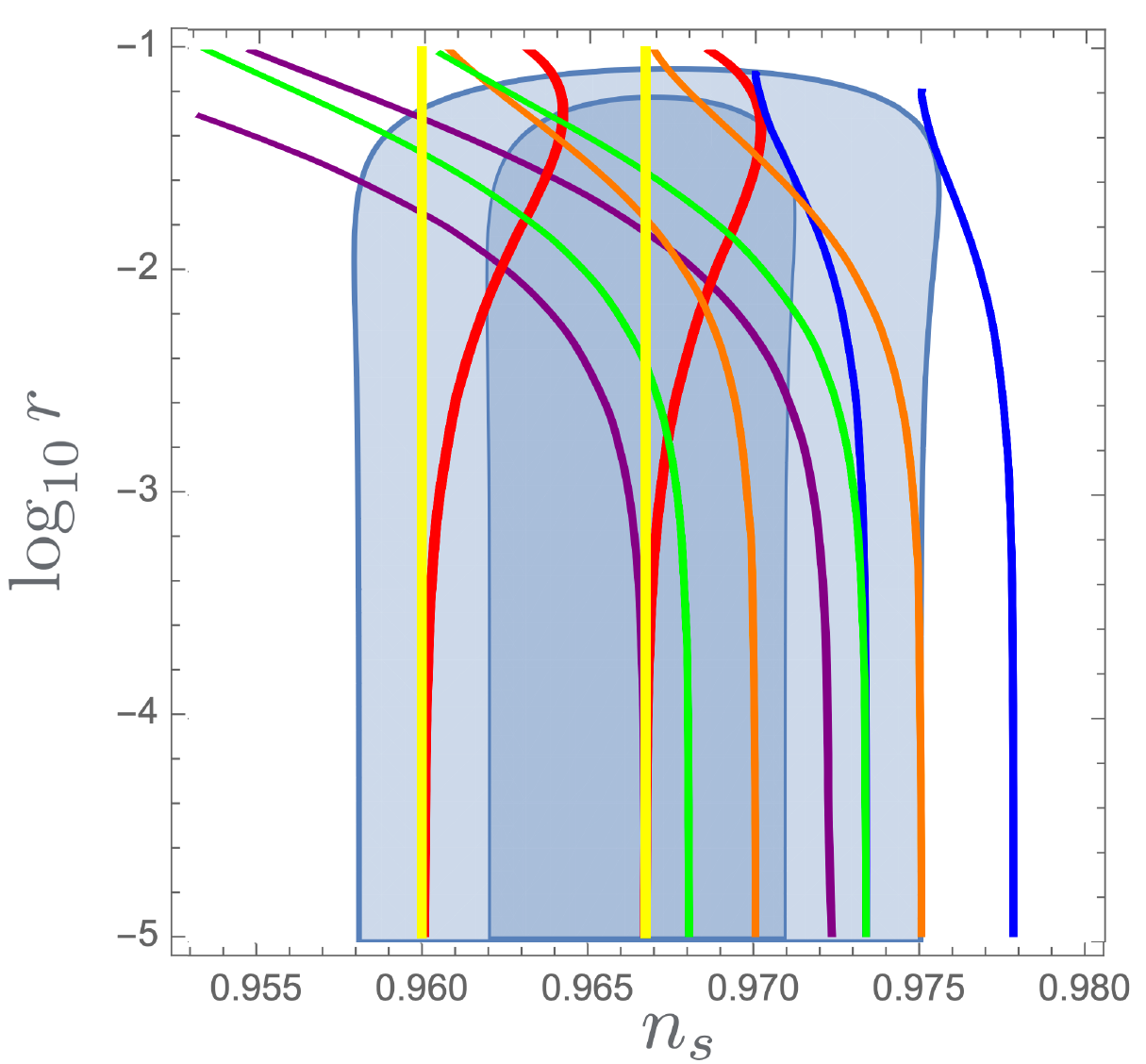}
\vskip -5pt
\caption{\footnotesize A combined plot of the predictions of the simplest exponential $\alpha$-attractor models and polynomial $\alpha$-attractor models for $N_* = 50$ and $60$ \cite{Kallosh:2019hzo}. From left to right, we show predictions of T-models and E-models (yellow and red lines) and of polynomial $\alpha$-attractors with $k=4,3,2,1$  (purple, green, orange, and blue lines).  These models, together, fully cover the blue area in this plot (corresponding to {\it Planck} 2018).}
\label{Blue}
\end{figure}

 But these results are valid only for a sufficiently small $r$. For more detailed information, consider Fig. \ref{Blue} where the blue data background corresponds to the {\it Planck} 2018 results, including BAO. A set of exponential $\a$-attractors  \cite{Kallosh:2013yoa} plus polynomial $\a$-attractors with $k=4,3,2,1$  \cite{Kallosh:2022feu} covers the entire blue area in this plot.  Using Fig. \ref{Blue} as our guide, we will discuss a possible interpretation of various values of $n_{s}$  and $r$.  

First of all, at $r \gtrsim  5\times 10^{-3}$, the predictions of all models except the T-models are far from their attractor regime. Therefore, finding, for example, that  $n_{s} \approx 0.965$ could be explained both in the context of exponential attractors, and polynomial attractors with $k = 4,3,2$.  Meanwhile for $r < 5\times 10^{-3}$, the result $n_{s} \approx 0.965$ would strongly favor the exponential attractors.

More importantly, the simplest exponential attractors are incompatible with the P-ACT-LB result $n_s=0.9739\pm 0.0034$, but this result is compatible with polynomial attractors with $k = 1,2$.  With increasing accuracy in determining $ n_s$, one could be able not only to distinguish between the exponential and polynomial attractors, but even specify the parameter $k$.  It may be important in this respect that  Simons Observatory plans to halve the current error bar on the scalar perturbation spectral index $n_{s}$ \cite{SimonsObservatory:2025wwn}.

A particularly interesting and instructive example is provided by the basic chaotic inflation model $\tfrac12 {m^{2}\phi^{2}}$ with nonminimal coupling to gravity $(1+ \phi)R$\,  \cite{Kallosh:2025rni}. For $N_* = 60$, it predicts $n_{s} = 0.974$, which is compatible with the P-ACT-LB constraint. The predicted value of $r$ is $0.0092$, which is almost 3 times higher than the predictions of the Starobinsky model and Higgs inflation for $N_* = 60$. Therefore, this model can be an interesting target for the B-mode searches at $n_s > 0.97$. In the Einstein frame, the non-minimal coupling disappears, and the potential of the canonically normalized inflaton field $\vp$ has a plateau. At large $\vp$ one has $V \approx \tfrac12 m^{2} (1-8/\vp^{2}+\dots)$, which is the potential with a power-law approach to the plateau, corresponding to $k = 2$.

We should note that in this section, we concentrated on the simplest single-field models. Once we step away from single-field models to multi-field inflationary models, many new possibilities emerge  \cite{Kallosh:2025ijd}. 
 
If the further investigations confirm that $n_s \sim  0.973 - 0.974$, it would indicate that {\it the data prefer potentials approaching the plateau polynomially rather than exponentially.}  This would change the narrative, which was very popular during the last decade, as represented by Fig. \ref{Flauger0}, where only the left part of the figure has targets for future B-mode experiments.  That is why it is so important to understand whether one can fully rely on the P-ACT-LB constraints on $n_{s}$, or the shift to larger $n_{s}$ is a consequence of combining current CMB and BAO data despite the significant tensions between these two datasets under the assumption of $\Lambda$CDM. Based on the results of this paper, we urge caution in interpreting these constraints until the BAO-CMB tension is resolved.

\vspace{1cm}

\section{Discussion}
\label{sec:discussion}

The scalar spectral index $n_s$ is a powerful probe of the primordial universe. 
Of the 6 $\Lambda$CDM parameters, $n_s$ is unique for its ability to directly constrain fundamental physics: while typical cold dark matter models can easily accept shifts in $\Omega_c$, and dark energy models can easily accommodate shifts in $\Omega_{\Lambda}$, inflationary models live or die (with some exceptions, as highlighted above) on $n_s$, and even percent-level shifts in $n_s$ are enough to reject some models in favor of others. 

In this work, we have endeavored to understand the constraints on $n_s$ that derive from the combination of CMB and BAO data, and in particular, CMB data from ACT DR6
and {\it Planck} 2018, and BAO data from DESI DR2. While ACT, SPT, and {\it Planck}, individually and in combination, are all compatible with the benchmark inflationary models targeted by CMB-S4, LiteBIRD, and other next-generation CMB experiments\cite{SPT-3G:2025vtb}, the combination of current CMB data with DESI leads to a shift in $n_s$ that disfavors these same models.

Through dedicated analyses of CMB data alone and in combination with DESI, we have demonstrated that the shift in $n_s$ when the two are combined is a product of the discrepancy in the BAO parameters $r_dh$ and $\Omega_m$ as inferred from DESI BAO data and from the CMB, which we term the {\it BAO-CMB tension}, and the tight correlation between $n_s$ and the BAO parameters in the fit of $\Lambda$CDM to CMB data.

The resolution of the BAO-CMB tension remains an open question, and so too does the fate of $n_s$. It may be that the BAO-CMB tension is resolved by future observations, or, if not, then the resolution may lie in new physics beyond $\Lambda$CDM. Neither possibility makes a firm prediction for whether $n_s$ will remain in the range currently favored by the CMB alone or else move to the relatively larger $n_s$ values of CMB+DESI (assuming $\Lambda$CDM). We have considered both possibilities in detail in Sec.~\ref{sec:inflation}. The latter scenario would prefer inflationary potentials approaching a plateau polynomially rather than exponentially, while the former would maintain the preference for the benchmark models of Higgs inflation, Starobinsky inflation, and (exponential) $\alpha$-attractors.

Future CMB data will play an important role, in particular in the determination of $n_s$. On-going analyses of SPT data covering a total of $10,000\,\mathrm{deg}^2$ ($f_{sky}\approx 25\%$) will greatly improve its ability to constrain $n_s$ and reach {\it Planck} precision \cite{spt_forecast}. Simons Observatory aims to halve the current error bar on $n_{s}$ \cite{SimonsObservatory:2025wwn}, reaching down to $\sigma(n_s)=0.002$.

Within $\Lambda$CDM, one can also bring the two probes into better agreement by raising $\tau_{\rm reio}$ to $\sim\,0.09$ \cite{Sailer:2025lxj,Jhaveri:2025neg}.
However, the large-angular-scale E-mode polarization measurement by {\it Planck} leaves little room for such a shift.
Still, improved measurements of this part of the CMB spectrum therefore remain interesting in the context of the BAO-CMB tension \cite{essinger14, watts18, li25, hazumi20, sakamoto22}.

On the other hand, the resolution may lie beyond $\Lambda$CDM; indeed, a number of extended models are able to better accommodate CMB and BAO at the same time. For example:  
spatial curvature, time-evolving dark energy, and a vanishing neutrino mass (see Table VII in \cite{SPT-3G:2025bzu}).
 However, it is important to bear in mind the details of the model one is looking at; for example, the required decrease of the sum of neutrino masses below $0.06\,\mathrm{eV}$ is in discord with oscillation experiments.
Importantly, the statistical evidence for these departures from the standard model remains moderate at the $2-3\,\sigma$ level.
Increasing the BAO-CMB tension in $\Lambda$CDM through improved data may lead to a stronger preference for new physics and may identify one model that clearly outperforms the others.

While this work has focused on DESI BAO, it is worth noting that other large-scale structure experiments can 
be informative as well, such as the Dark Energy Survey (DES) BAO \cite{DES:2024pwq}, galaxy clustering and weak lensing \cite{DES:2021wwk}, and supernovae \cite{DES:2018paw}. This would mirror the Hubble tension, where including varied large-scale structure datasets significantly strengthens the constraints on resolutions based on physics beyond $\Lambda$CDM \cite{Hill:2020osr,Ivanov:2020ril,Herold:2021ksg,Herold:2022iib,Goldstein:2023gnw,McDonough:2023qcu}. Looking towards the future, the Euclid experiment aims to dramatically increase the precision of BAO measurements \cite{Euclid:2025dlg}. The Vera Rubin Observatory will also be able to constrain the evolving dark energy models that can possibly address the BAO-CMB tension, through a variety of probes \cite{LSSTScience:2009jmu}.

There are thus many directions for future work, such as further development of beyond-$\Lambda$CDM models to resolve the BAO-CMB tension, the comparison with other cosmological data sets and dataset combinations, such as other LSS data (like BOSS full shape, DES), SN data, and the interplay with other cosmological parameter tensions such as the Hubble tension. We leave these and other possibilities to future work.


\acknowledgments

The authors thank  J.R.~Bond, M.~Cort\^es, G.~Efstathiou, F.~Finelli, W.~Freedman, J.~C.~Hill, A.~Liddle, D.~Roest, L.~Thiele, Y.~Yamada and J.~Yokoyama for helpful discussions. E.M. is supported in part by a Discovery Grant from the Natural Sciences and Engineering Research Council of Canada and by a New Investigator Operating Grant from Research Manitoba. E.M. thanks Kavli IPMU for hospitality while a portion of this work was completed. Kavli IPMU is supported by the World Premier International Research Center Initiative (WPI), MEXT, Japan.
EGMF thanks the support of the Serrapilheira Institute. L.K. is supported in part by DOE Office of Science award DESC0009999 and the Michael and Ester Vaida Endowed Chair in Cosmology and Astrophysics. L.K. also thanks Kavli IPMU for hosting him during an early conversation with E.M. and EGMF. that eventually led to this work.
L.B. has received funding from the European Research Council (ERC) under the European Union’s Horizon 2020 research and innovation program (grant agreement No 101001897), and funding from the Centre National d’Etudes Spatiales.  R.K and A.L. are supported by SITP and by NSF Grant PHY-2310429. \\

\bibliography{refs}

\end{document}